\documentclass[conference]{IEEEtran}
\makeatletter
\def\ps@headings{%
\def\@oddhead{\mbox{}\scriptsize\rightmark \hfil \thepage}%
\def\@evenhead{\scriptsize\thepage \hfil \leftmark\mbox{}}%
\def\@oddfoot{}%
\def\@evenfoot{}}
\makeatother
\usepackage{setspace}
\usepackage{bm}
\usepackage{cite}
\usepackage{lettrine}
\usepackage{amssymb}
\usepackage{amsmath}
\usepackage{amsfonts}
\usepackage{enumerate}
\usepackage{enumitem}
\usepackage{indentfirst}
\usepackage{graphicx}
\usepackage{epstopdf}
\usepackage{multirow} 
\usepackage{makecell}
\usepackage{xcolor}
\usepackage{listings}
\usepackage{float}
\usepackage{subfigure}
\usepackage{amsthm}
\usepackage[top=0.75in, bottom=1in, left=0.65in, right=0.65in]{geometry}
\usepackage{algpseudocode}
\usepackage{algorithmicx,algorithm}
\usepackage{algorithm}  
\usepackage{breqn}
\newcommand{\mm}{Meta-IoT }
\newcommand{\beq}{\begin{equation}} 
\newcommand{\eeq}{\end{equation}}

\begin{document}
\title{\huge{Deployment Optimization for Meta-material Based \\Internet of Things}}

\author{
\IEEEauthorblockN{
\small{Xu~Liu}\IEEEauthorrefmark{1},
\small{Jingzhi~Hu}\IEEEauthorrefmark{2},
\small{Hongliang~Zhang}\IEEEauthorrefmark{3},
\small{Boya~Di}\IEEEauthorrefmark{2},
\small{and Lingyang~Song}\IEEEauthorrefmark{2},\\}
\IEEEauthorblockA{
 	\IEEEauthorrefmark{1}\small{College of Engineering, Peking University, Beijing, China,}\\
    \IEEEauthorrefmark{2}\small{Department of Electronics, Peking University, Beijing, China,}\\
    \IEEEauthorrefmark{3}\small{Department of Electrical Engineering, Princeton University, Princeton, NJ, USA}\\}
}

\maketitle

\setlength{\abovecaptionskip}{0pt}
\setlength{\belowcaptionskip}{-10pt}

\begin{abstract}
In this paper, we propose a \mm system to achieve ubiquitous deployment and pervasive sensing for future Internet of Things (IoT).
In such a system, sensors are composed of dedicated meta-materials whose frequency response of wireless signal is sensitive to environmental conditions. Therefore, we can obtain sensing results from reflected signals through \mm devices and the energy supplies for IoT devices can be removed.
Nevertheless, in the \mm system, because the positions of the \mm devices decide the interference among the reflected signals, which may make the sensing results of different positions hard to be distinguished and the estimation function should integrate the results to reconstruct 3D distribution. It is a challenge to optimize the positions of the \mm devices to ensure sensing accuracy of 3D environmental conditions.
To handle this challenge, we establish a mathematical model of \mm devices' sensing and transmission to calculate the interference between \mm devices.
Then, an algorithm is proposed to jointly minimize the interference and reconstruction error by optimizing the \mm devices' position and the estimation function.
The simulation results verify that the proposed system can obtain a 3D environmental conditions' distribution with high accuracy.
\end{abstract}


\section{Introduction}
In the upcoming 6G systems, Internet of Things (IoT) plays an important role in various applications, such as intelligent industrial manufacturing and environmental protection
\cite{8624565}. As a result, an extremely large number of IoT sensors are needed to collect sensory data of the environment, which will be approximately 10-fold more than that in 5G~\cite{Liu2020Vision}.
However, the existing sensors face challenges to support the dense deployment of IoT devices, as they need power supplies for sensing and transmission. Therefore, traditional sensors cannot  continuously work without any human intervention for an ultra-long time~\cite{zhang2020beyond}.
To achieve pervasive environment sensing, it is necessary to develop sensors working without power supplies.

Fortunately, meta-material shows the potential to work as passive sensors satisfying the no-battery requirement of 6G sensing applications, which we refer to as \mm devices~\cite{CO2}. The \mm devices are printed circuits on supportive substrates combined with some sensitive materials and work as the passive wireless signal reflector. Therefore, their frequency response for wireless signals can be exploited to be sensitive to specific target. Compared with active wireless sensors, \mm devices are capable of simultaneous sensing and transmission and have no power supplies and maintenance, negligible volume, and ultra-low-cost.
 
In this paper, we consider the \mm system to obtain the 3D distribution of environmental conditions. In such a system, several \mm devices are deployed and a pair of the wireless transceiver is used to measure the reflected signals. Then, the distribution of environmental conditions is estimated by the signals reflected by \mm devices.
Nevertheless, it remains a challenge to reconstruct the 3D distribution of environmental conditions by using the \mm system. This is because the \mm device worked as the passive wireless signal reflector and the reflected signals from \mm devices may interfere with each other, which makes the sensing results of different positions hard to be distinguished leading to high sensing errors through these signals.

To address this issue, we first analyze the transmission model between the transceiver. Then, according to the transmission model, we further analyze how the \mm devices' position influences the interference among these \mm devices.
In order to handle the interference among \mm devices and minimize the reconstruction error, we formulate a joint \mm devices' position and estimation function optimization problem.
As the formulated problem is NP-hard, we decompose it into two sub-problems: interference minimization problem and estimation function optimization problem, and solve them sequentially. 
By using numerical simulation, we verify the capability of the proposed system to reconstruct the 3D distribution of multiple environmental conditions with low error.

In the literature, several works have discussed the use of the meta-material for various types of sensing applications, such as gas concentration~\cite{CO2}, humidity~\cite{ekmekci2019use}, strain~\cite{melik2010metamaterial} and so on.
However, these works focus on the design to improve the sensing performance while the transmission is neglected. Moreover, only a single \mm is deployed in the system, which can only obtain the results of a specific area. In our system, to achieve 3D distribution of environmental conditions sensing,  several \mm devices are deployed and the mutual interference in the transmission process is also considered.

The rest of this paper is organized as follows. In Section~\ref{sec: mm sensor}, we present the design of the \mm device. In Section~\ref{sec: system model}, we introduce the \mm system and analyze the transmission model. Then, in Sections~\ref{sec: problem formulation} and~\ref{sec: algorithm}, to minimize the reconstruction error of the distribution of environmental conditions, we formulate a joint \mm devices' positions and estimation function optimization problems and solve them, respectively. In Section~\ref{sec: results}, we demonstrate the simulation results. Finally, conclusions are drawn in Section~\ref{sec: conclusion}.
	
\section{Meta-IoT Devices}
\label{sec: mm sensor}
In this section, we first describe the components and the structure of a \mm device unit. Then, we proposed an equivalent circuit model and establish the model of the reflection coefficient of the \mm device based on the equivalent circuit model.

\begin{figure}[!t] 
\center{\includegraphics[width=0.73\linewidth]{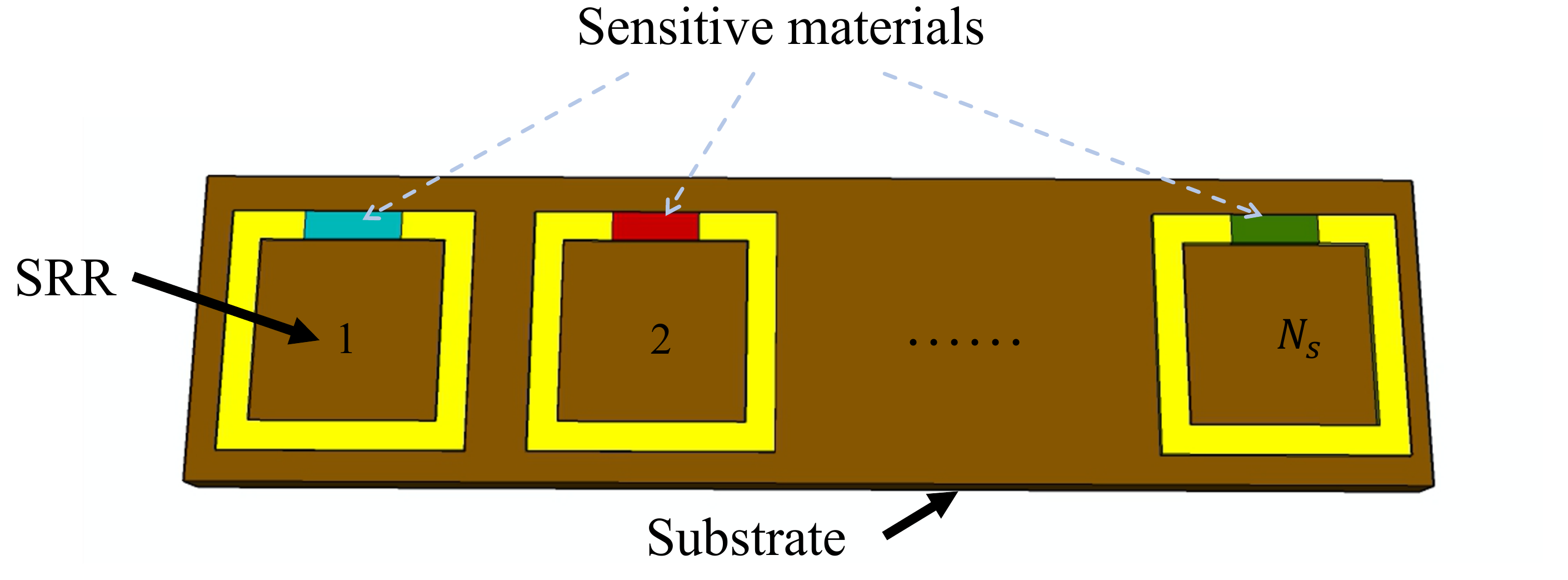}}
	\vspace{-0.5em}
	\caption{Layout of the \mm device unit. The \mm device consists of $N_s$ SRRs, which are labeled with their respective indexes.}
	\label{fig: mm sensor}
	\vspace{-0.3em}
\end{figure}

\subsection{Meta-IoT Unit}
\label{ssec: meta-IoT unit}
Meta-IoT devices are wireless passive devices constitute of meta-material.
By designing the component meta-material to have specific structures and sensitive materials, we can make the frequency response of the \mm device sensitive to the multiple environmental conditions.

As shown in~Fig.~\ref{fig: mm sensor}, the proposed \mm device unit is composed of $N_s$ split ring resonators (SRRs). Each SRR consists of a dielectric substrate and a square metal ring which is printed on the substrate and has a gap that is filled with sensitive materials. With different sensitive materials, each SRR is designed to sense a certain environmental condition. Except for the sensitive material and the gap width, the $N_s$ SRRs have the same structure and are composed of the same material. Therefore, we take the $n$-th $(n \in [1,N_s])$ SRR as an example in the following to analyze the reflection coefficient. 

\subsection{Equivalent Circuit Model}
\label{ssec: circuit model}
\begin{figure}[!t]       
\center{\includegraphics[width=0.85\linewidth]{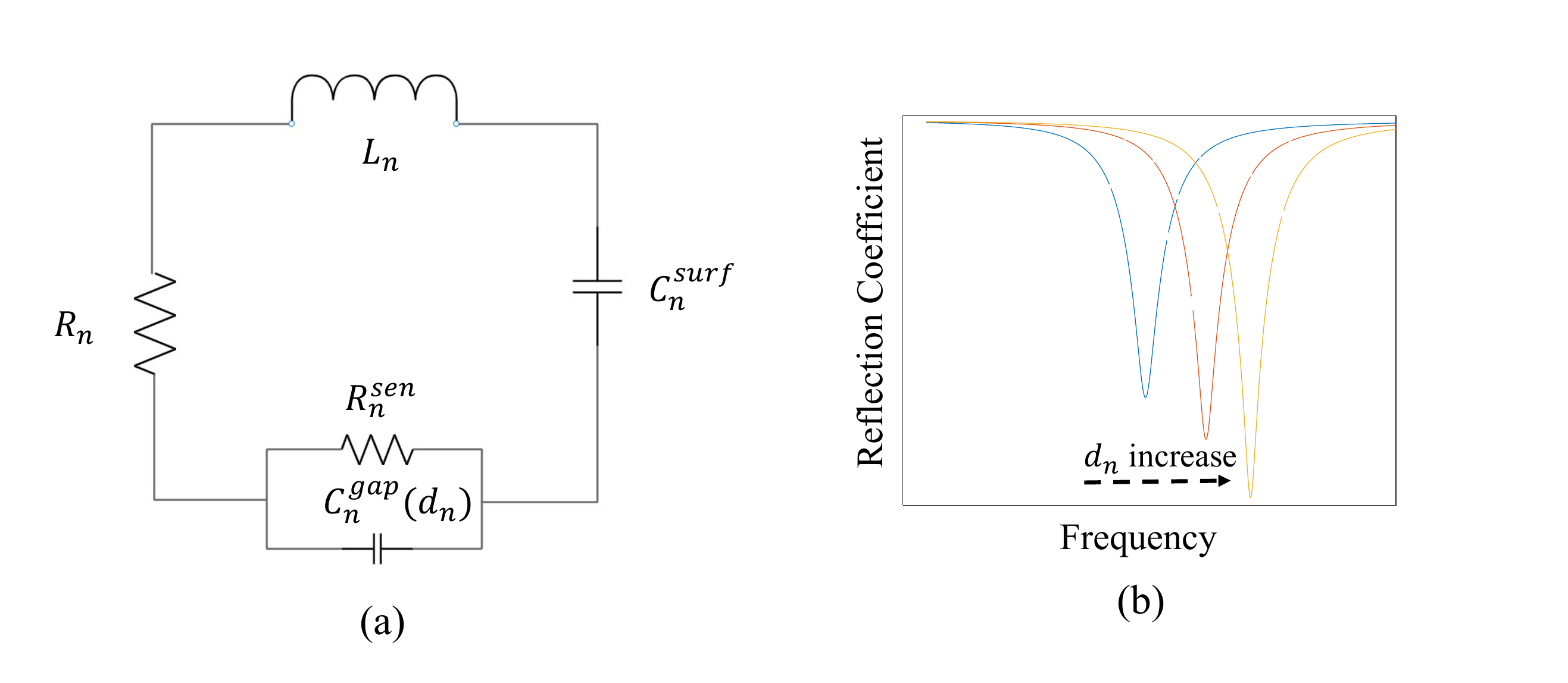}}
\vspace{-1em} 	
\caption{The circuit model for the sensing SSR, and the reflection coefficient with different $d_n$} 	
\label{fig: circuit_model_frequence_response}  
\vspace{-0.5em}
\end{figure}

Based on~\cite{microfluidicSRR}, the SRR can be approximated by an RLC resonant circuit as illustrated in Fig.~\ref{fig: circuit_model_frequence_response}~(a). For the $n$-th SRR with gap width $d_n$, the $n$-th aimed environmental condition being $c_n$ and the other environmental conditions being $\bm c_{-n}$.
Then, according to the circuit model given in Fig.~\ref{fig: circuit_model_frequence_response}~(a), the impedance can be calculated as
\beq
\setlength{\abovedisplayskip}{2pt}
\begin{aligned}
\label{equ: Z}
& Z_n(f, c_n,\bm c_{-n}, d_n)= R_n + 2\pi jfL_n + \frac{1}{2\pi jfC_{n}^{surf}}\\ 
&+ \frac{R_{n}^{sen}(c_n,\bm c_{-n}, d_n)}{ 1 + 2\pi jfC_{n}^{gap}(d_n)\cdot R_{n}^{sen}(c_n,\bm c_{-n}, d_n)},
\end{aligned}
\eeq
where $f$ denotes the frequency of wireless signals, $R_n$ donates the resistance of the metal ring, $L_n$ denotes the self-inductance, $C_{n}^{surf}$ denotes the capacity corresponding to the surface, $R_{n}^{sen}(c_n,\bm c_{-n}, d_n)$ denotes the resistance of the sensitive material at environmental conditions $c_n$ and $\bm c_{-n}$, and $C_{n}^{gap}(d_n)$ denotes the capacity caused by the gap whose width is $d_n$. 

\subsection{Reflection Coefficient Model}
\label{ssec: reflection coefficient}
Based on~(\ref{equ: Z}), the reflection coefficient of the signal SRR can be analyzed as follows. According to~\cite{TheTheoryOfElectromagnetism}, when the applied voltage on an SRR is $U$, the power consumption of the circuit can be calculated by
\beq
\setlength{\abovedisplayskip}{2pt}
\setlength{\belowdisplayskip}{4pt}
\label{equ: Pconsume}
p_{n}^{con} = \frac{U^2}{|Z_n|^2}Re(Z_n),
\eeq
where $Re(\cdot)$ denotes the real part of a complex value. Besides, when $|Z_n|$ is at its minimum in terms of $f$, the circuit is \emph{resonance} and the frequency, i.e, $f_r$, is called the \emph{resonance frequency}. At the resonance of the circle, the power consumption of the circuit is at the maximum so that the power of reflected signals is at the minimum.
Therefore, there is an \emph{absorption peak} at the resonance frequency for the reflection coefficient, which is illustrated in Fig.~\ref{fig: circuit_model_frequence_response}~(b).

Based on~(\ref{equ: Z}) and (\ref{equ: Pconsume}), the sensitive material's resistance $R_{n}^{sen}$ influences the impedance $Z_n$, and thus influences the position of the absorption peak. Then, we can get the environmental condition by analyzing the position of the absorption peak. Especially, to depict the SRR's reflection coefficient, the power of the reflected signal can be modeled as
\beq
\setlength{\abovedisplayskip}{2pt}
\setlength{\belowdisplayskip}{2pt}
\label{equ: Preflect}
p_{n}^{ref} = p_{n}^{inc}-p_{n}^{con}.
\eeq
where $p_{n}^{inc}$ is the power of the incident signal, and $p_{n}^{ref}$ donates the power reflected by the SRR. We assume that the incident signal on the SRR to be a planer horizontal polarized electromagnetic wave with filed intensity $E$. Based on~\cite{TheTheoryOfElectromagnetism}, the incident signal power is proportional to $E^2$ and the parameter $U$ of $p_{n}^{con}$ depicts the electric potential difference which is proportional to $E$. Then, we have the function of the reflection coefficient for the single SRR as
\beq
\setlength{\abovedisplayskip}{2pt}
\setlength{\belowdisplayskip}{2pt}
\label{equ: SRRreflection}
s_{n}(f, d_n, c_n,\bm c_{-n})=\frac{p_{n}^{ref}}{p_{n}^{inc}} = 1-a\frac{Re(Z_n)}{|Z_n|^2}.
\eeq

Due to there are multiple SRRs involved in the \mm device unit, to distinguish the absorption peak cased by different SRRs, we need to design the different structural parameters for each SRR.
As the resonance frequency increases with gap width $d_n$, the absorption peak moves to the right when $d_n$ increases in Fig.~\ref{fig: circuit_model_frequence_response}~(b). Therefore, we are able to design SRRs with different gap widths to make their absorption peak in different frequency points. Then, based on~(\ref{equ: Preflect}) and (\ref{equ: SRRreflection}), the reflection coefficient of the \mm device unit can be expressed as
\beq
\setlength{\abovedisplayskip}{2pt}
\setlength{\belowdisplayskip}{2pt}
\label{equ: devicereflection}
S(f, \bm d, \bm c)=\frac{1}{N_s}\sum_{i\in [1,N_s]}{s_{i}},
\eeq
where $\bm d$ donate the gap widths of $N_s$ SRR. Besides, the reflection coefficient of the \mm device plays an important role in describing device performance. However, to obtain a precise reflection coefficient, numerical finite-element simulation, such as CST software~\cite{CST}, and practical experiments are in need.

\section{System Model}
\label{sec: system model}
\begin{figure}[!t] 
    \center{\includegraphics[width=0.85\linewidth]{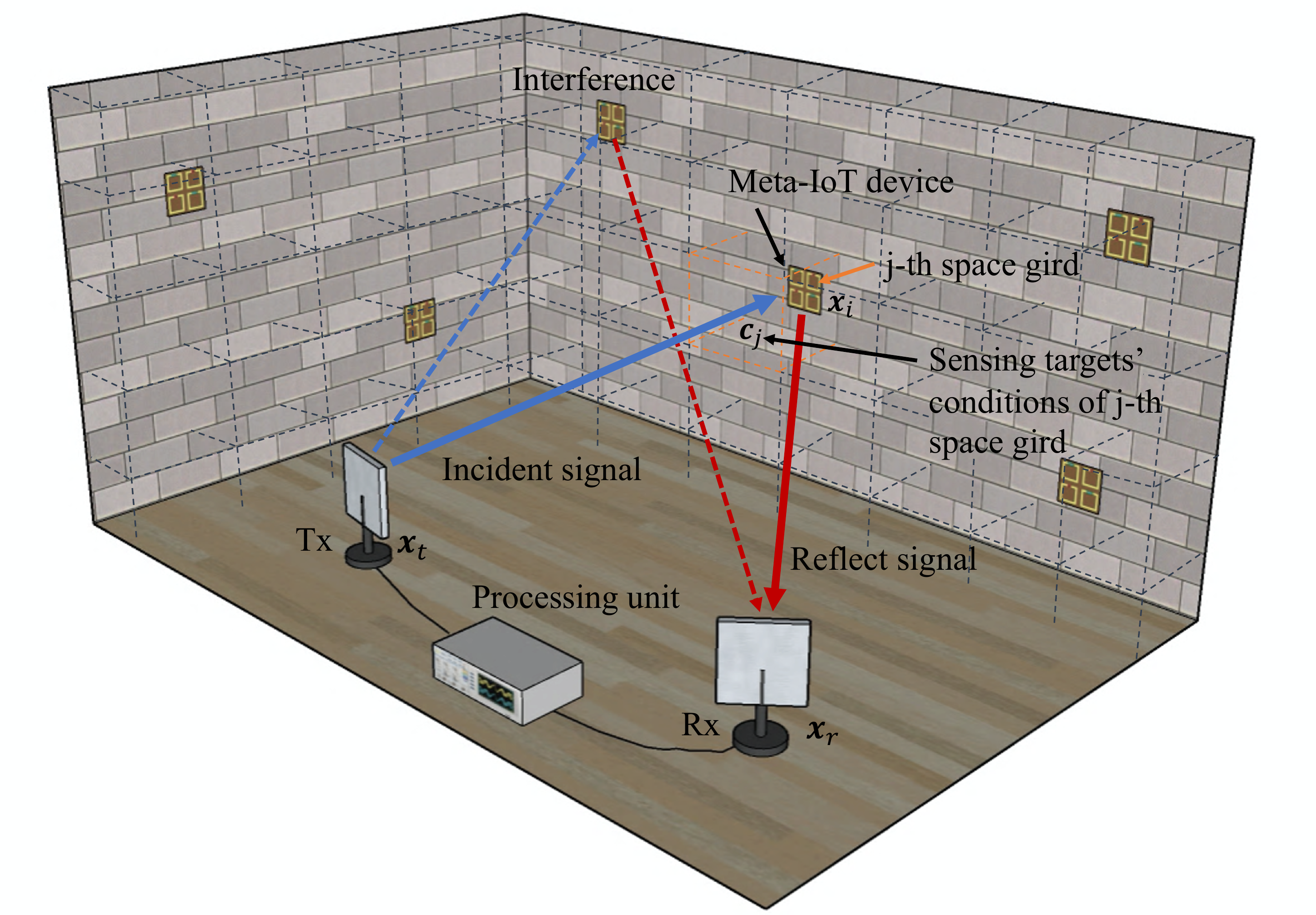}} 
	\caption{\mm high-resolution wireless sensing system for monitoring the indoor environmental conditions}
	\label{fig: system scenario}
\end{figure}

In this section, we first describe the proposed \mm system to reconstruct the 3D distribution of multiple environmental conditions. Then, the transmission model of the wireless signals in this system is established. Following that, we introduce the \mm sensing protocol for taking the measurements. 

\subsection{System Description}
\label{ssec: system}
As illustrated in Fig.~\ref{fig: system scenario}, the system consists of a pair of transceivers, which is equipped with a processing unit, a Tx and an Rx antenna array, and $N$ \mm devices. Each \mm device is composed of $N_u \times N_u$ units densely paved as a 2-dimensional rectangle. Besides, by controlling the imposed phase shifts on different antennas, the Tx and the Rx antenna arrays are capable of steering their beams towards different directions, so that they can transmit polarized directional signals with bandwidth $[f_l, f_u]$ to each \mm device and receive the reflected signal. 

The positions of the Tx and Rx antenna arrays can be expressed as $\bm x_{t}$ and $\bm x_{r}$. Besides, the $N$ \mm devices are deployed on the walls around the room, as illustrated in Fig.~\ref{fig: system scenario}. We denote the available position set where the \mm device can be deployed as $\mathcal S_{X}$. Here, the position of the $i$-th \mm device is denoted by the $\bm x_{i}, (\bm x_{i}\in \mathcal  S_{A}, \forall i \in [1, N])$ and the positions of the $N$ \mm devices constitute a set~$\mathcal X$. 

As described in Section~\ref{sec: mm sensor}, the \mm devices' reflection coefficients are sensitive to the environmental conditions. To sense the distributions of environmental conditions, the Tx antenna array sends a polarized wireless signal to each device, and the Rx antenna array receives the signals reflected by the \mm device.

The received signal powers are influenced by the reflection coefficients of \mm devices, which are determined by the environmental conditions. Therefore, the processing unit can potentially obtain the sensing results by analyzing the received signals' power. Moreover, the processing unit adopt an \emph{estimation function} to reconstruct the distribution within a \emph{target space} around the \mm device. Without loss of generality, we assume that the target space is discretized into $M$ space grids, and denote the set of $M$ space grids as $\mathcal M$. The detail of the estimation function will be discussed in Section~\ref{sec: algorithm}.

\subsection{Transmission Model}
\label{ssec: transmission model}
In such a system, the Tx and Rx arrays have line-of-sight~(LOS) paths to each \mm device. Without loss of generality, we take the case when the Tx array sends directional signals towards the $i$-th \mm device as an example to show the signal transmission model.

Based on the Friis' Free Space Link Model in~\cite{Friis}, the power of the incident signal to the $i$-th \mm device can be calculated by
\beq
\label{equ: Sis}
P_{i}^{inc}(f) = P_{T}\cdot \sigma \cdot\frac{1}{4 \pi r_{i,t}^2}G_i^t(\theta_i^t, \varphi_i^t, f),
\eeq
where $P_{T}$ denotes transmit power, $\sigma$ denotes the area of a \mm device, $f$ is the frequency of the incident signal, and $r_{i,t} = \|\bm x_{i} - \bm x_t \|_2$ is the distance between $i$-th device and Tx. Besides, $G_i^t(\theta_i^t, \varphi_i^t, f)$ is the gain factor of the Tx antenna array when measuring the $i$-th device, and $(\theta_i^t, \varphi_i^t)$ is the azimuth angle vector of the $i$-th device relative to the Tx.

At the Rx antenna array, the received power of the signals reflected by the $i$-th device can be expressed as
\beq
\label{equ: Pr,i}
P_i^{rec}(\bm c, f) = P_i^{inc}(f)\cdot S_i(\bm c, f) \cdot \frac{1}{2 \pi r_{i, r}^2} \frac{G_i^r(\theta_i^r, \varphi_i^r, f) {\lambda}^2}{4 \pi},
\eeq
where $S_i$ indicates the reflection coefficient of the $i$-th \mm device defined in~(\ref{equ: devicereflection}), $\lambda$ is the wavelength of the incident signal and $r_{i,r}$ is the distance between $i$-th device and Rx antenna array. Besides, $ G_i^r(\theta_i^r, \varphi_i^r, f)$ is the gain factor of the Rx antenna when measuring the $i$-th device and $(\theta_i^r, \varphi_i^r)$ is the azimuth angle vector of the $i$-th \mm device relative to the Rx.

Based on~(\ref{equ: Sis}) and~(\ref{equ: Pr,i}), when measuring the i-th \mm device, the total received signal power can be expressed as follows, which consists of four parts, i.e., the \emph{target}, the \emph{interference}, the \emph{environmental reflection}, and the \emph{noise}, which can be expressed as
\beq
\begin{aligned}
\label{equ: P_R}
&P_{R,i}(\bm c, f)[dB]=10\log_{10}(\\
&\underbrace{\frac{P_{T}\sigma \lambda^2}{32\pi^3 r_{i,r}^2r_{i,t}^2}\cdot S_{i}(\bm c, f) G_i^t({\theta}_i^t, {\varphi}_i^t, f) G_i^r({\theta}_i^r, {\varphi}_i^r, f)}_{\text{Target}}+\\
&\underbrace{\sum_{j\in[1,N]}^{j\neq i}{\frac{P_{T}\sigma \lambda^2}{32\pi^3 r_{j,r}^2r_{j,t}^2}S_{j}(\bm c, f) G_i^t({\theta}_j^t, {\varphi}_j^t, f) G_i^r({\theta}_j^r, {\varphi}_j^r, f)}}_{\text{Interference}}\\
&+\underbrace{\eta P_{T}\cdot R_{env}}_{\text{Environmental reflection}})+\underbrace{\varepsilon(f)}_{\text{Noise}}.
\end{aligned}
\eeq

\subsubsection{Target} The first part of received power is carried by the signals that are reflected by the \mm device to be sensed.
\subsubsection{Interference} The second part is due to the signals reflected by the \mm devices other than the target device.
\subsubsection{Environmental reflection} The third part is the power reflected by the surrounding environment, where $R_{env}$ donates the reflection coefficient and can be obtained with the help of~\cite{wallreflect}, $\eta$ is the ratio of the transmitted power reflected by the environments and total transmitted power.
\subsubsection{Noise} The fourth part donates the noise, i.e., $\varepsilon$, in the power measurement, which is assumed to follow a Gaussian distribution $\mathcal(0, \sigma_m)$, with $\sigma_m$ being the variance of the measurement noise.

Besides, in~(\ref{equ: P_R}), it can be observed that the received signal power is influenced by the $S_{i}(\bm c, f)$. If $S_{i}(\bm c, f)$ can be derived from the received power, the environmental conditions at the $i$-th \mm device can be estimated. Nevertheless, as can be observed in~(\ref{equ: P_R}), suffers from the interference due to the signals reflected by other \mm devices as well as the environment scattering, which makes $S_{i}(\bm c, f)$ hard to estimate. To handle this issue, we propose a novel method of deriving distribution of environmental condition from the received signals, which will be described in detail in Section~\ref{sec: problem formulation}.

\subsection{Meta-sensing Protocol}
\label{ssec: protocol}
To obtain the distribution of environmental conditions through the received signals' power, we propose the following \mm sensing protocol to coordinate the measuring process of the $N$ \mm devices’ reflected signals at frequencies $f_1$, ..., $f_L$, where $\{f_i\}_{i=1}^L$ are a set of discretely sampled frequencies within spectrum band $[f_{l}, f_{u}]$. The protocol is described as follows.

The measuring process is carried periodically, and in each measuring period, the $N$ \mm devices are measured sequentially.
Specifically, in the $i$-th measurement $(i\in[1,N])$, the Tx and Rx antenna arrays steer their main-lobes towards the $i$-th \mm device.
Then, the Tx sends the wireless signals at frequencies $f_1$, ..., $f_{L}$, sequentially, and the Rx respective record the received signal power values. Thus, for the $i$-th device's wireless signal, the received signal powers can be expressed by an $L$ dimensional column vector, i.e., $\bm p_{R,i}$. For $N$ \mm devices, the received signal powers can be expressed as an $L \times N$-dimensional matrix, i.e., $\bm P_{R}=[\bm p_{R,1},...,\bm p_{R,N}]$. We refer to $P_R$ as the \emph{measurement matrix}.

At the end of each period, the obtained measurement matrix, i.e., $\bm P_R$, is sent to the processing unit, which handles it by the estimation function and obtains the estimation of the $N_s$ environmental condition distribution in the current period.

\section{Problem Formulation}
\label{sec: problem formulation}

In this section, a \emph{joint position and estimation function optimization problem} is formulated to reconstruct a 3D distribution of environmental conditions with minimal error. 

To begin with, we first define the estimation function properly, which is used by the processing unit to estimate the distributions of the environmental conditions based on the measurement matrix, i.e., $\bm P_{R}$. We has discretized the target space into $M$ space grids, and the $N_s$ environmental conditions within the $m$-th space grid is denoted by a column vector, $\widetilde{\bm c}_{m}$, Then, the distribution can be expressed as an matrix, i.e., $\widetilde{\bm C}=[\widetilde{\bm c}_{1},...,\widetilde{\bm c}_{M}]$. Thus, the estimation function can be expressed as a mapping from $\bm P_{R}$ to $\widetilde{\bm C}$. Without loss of generality, we assume that the estimation function is a parametric function $\bm f^{\bm w}: \bm P_{R} \rightarrow \widetilde{\bm C}$, with $\bm w$ being the parameter vector.

The objective is to minimize the difference between the estimated environmental condition and the true values in the target space $\mathcal M$, which we refer to as the \emph{reconstruction error}.
From the system model, it can observe that the reconstruction error is influenced by two factors, which are the \mm devices' position and the estimation function of the processing unit. First, the \mm devices' position influences the interference between each \mm device, which makes reflection coefficient, i.e., $S_i(\bm{c}, f)$, hard to estimate. Then, the estimation function determines the ability of the processing unit to reduce the \mm reflection coefficient in the received signal power matrix, as well as the ability to reconstruct the distributions of the environmental conditions by using the \mm reflection coefficient.

Therefore, we need to minimize the reconstruction error by optimizing the \mm devices' position, i.e., $\mathcal X$ and the estimation function parameters, i.e., $\bm w$. Then, the problem can be formulated as 
\begin{subequations}
\begin{alignat}{2}
\label{P1}
(P1): &\min_{\bm w, \mathcal X}\ &&L_{RMSE}(\bm w, \mathcal X)=\sum_{\bm c_{i}\in \mathcal C}{\|\widetilde{\bm c_{i}}- \bm c_{i}\|^2},\\
\label{P1: C1}
&s.t. &&\widetilde{\bm C} = \bm f^{\bm w}(\bm P_{R}),\\
\label{P1: C2}
& && \mathcal X \subseteq \mathcal S_X,
\end{alignat}
\end{subequations}
where $\mathcal C$ is the set of the known distributions of environmental conditions in the $M$ space grids. $\bm c_i$ means the environmental conditions of $i$-th space grid. Here, in~(\ref{P1}), $L_{RMSE}(\bm w, \mathcal X)$ denotes the sum of root mean squared error~(RMSE) given parameters $\bm w$ and $\mathcal X$. Constraint~(\ref{P1: C1}) indicates that the estimated environmental conditions are obtained by the transceiver using the estimation function with parameter $\bm w$ and measurement matrix $\bm P_{R}$. Besides, constraint~(\ref{P1: C2}) is due to the devices' position must be in accord with the available position set. 

\section{Algorithm Design}
\label{sec: algorithm}
Since the constraints are non-convex and the estimation function is non-linear, $(P1)$ is an NP-hard problem. Besides, the function parameter $\bm w$ is coupled with the devices' positions, which makes problem $(P1)$ even harder to solve.

To handle these issues, we decompose $(P1)$ into two sub-problems and propose the algorithms to solve them sequentially. First, an \emph{interference minimization problem} is proposed to minimize the interference through \mm devices' position optimization. Then, an \emph{estimation function optimization problem} is aimed to minimize the reconstruction error through optimizing the estimation function.

\subsection{Interference Minimization Problem}
\label{ssec: sP2}
The purpose of the interference minimization problem is to guarantee that the mutual interference between different \mm devices is as low as possible. 
The optimization variables $\mathcal X$ are the device position parameters, i.e., $\bm x_{i}$. Base on the transmission model proposed in~(\ref{equ: P_R}), the interference comes from the second part of the received signal power. Then, we aim to minimize the max interference by maximizing the minimal ratio between the channel gains of the target signals and the interference signals. 

In the measurement of the i-th \mm device, we denote the channel gain of each device reflected signal by $g_{i,j}$. Therefore, based on~(\ref{equ: P_R}) the interference minimization problem can be formulated as:
\begin{subequations}
\begin{alignat}{2}
\label{sP2}
(sP2): & \max_{\mathcal X} &&\min_{i\in[1,N]} \frac{g_{i,i}(\bm x_i)}{\sum_{j\in[1,N]}^{j\neq i} g_{i,j}(\bm x_i, \bm x_j)},\\
\label{sP2: C1}
&s.t. &&(\ref{P1: C2}),\\
\label{sP2: C2}
&     && g_{i,j}=G_i^t({\theta}_j^t, {\varphi}_j^t, f) \cdot G_i^r({\theta}_j^r, {\varphi}_j^r, f),\\
\label{sP2: C3}
&     &&
\begin{aligned}
(\theta_{j}^r,\varphi_{j}^r) = H(\bm x_{j}, \bm x_r),\\
(\theta_{j}^t,\varphi_{j}^t) = H(\bm x_{j}, \bm x_t),
\end{aligned}
\end{alignat}
\end{subequations}
which indicates that the signal power concentrate on only one device for each sensing action. Besides, based on geometric, the function $H$ in~(\ref{sP2: C3}) is able to be explicitly represented. 

To solve (sP2) efficiently, we adopt the simulated annealing algorithm, which can handle large combinatorial global optimization problems and avoid falling into local optimal, so that it has a high probability of finding the global optimal~\cite{vanLaarhoven1987}. The simulated annealing algorithm is based on the principle of changing the solving state to a worse value by probability which is decreased with iteration, so that it is capable to jump out the local optimal state. The algorithm terminates until there is no better result has been solved in several iterations or reaching the maximum number of iterations. 

Specifically, based on the $\mathcal S_X$ in~(\ref{sP2: C1}), the constraint of $(\theta_{j}^r,\varphi_{j}^r)$ and $(\theta_{j}^t,\varphi_{j}^t)$ can be calculated with the function~$H$. Then, $(\theta_{j},\varphi_{j})$ is optimized by the simulated annealing algorithm. Besides, we denote the resulting optimized \mm position set as $\mathcal X^*$, which will be used in the next estimation function optimization problem.

\subsection{Estimation Function Optimization Problem}
\label{ssec: sP3}
After solving the~(\ref{sP2}), the parameters $\mathcal X$ in~(\ref{P1}) can be regarded as the constant and we adopt the optimized \mm position set, i.e. $\mathcal X^*$. Then, the optimization variables are the estimation function parameters, i.e., $\bm w$, and the estimation function optimization problem can be formulated as:
\beq
\label{sP3}
\setlength{\abovedisplayskip}{4pt}
\setlength{\belowdisplayskip}{4pt}
(sP3): \min_{\bm w}\ L_{RMSE}(\bm w, \mathcal X^*), \qquad s.t.(\ref{P1: C1}).
\eeq	

To solve (sP3), we model that $f^w$ as a deconvolution neural network, which is an efficient model for mapping low-dimensional features to higher-dimensional features with deconvolution layers~\cite{dumoulin2016guide}. 
The neural network consists of a fully-connection layer, a deconvolution layer and two convolution layers, and uses the measurement matrix as input to calculate $\widetilde{\bm C}$. In this case, the parameter vector of the estimation function stands for the weights and the biases of each layer. 

Then, to obtain the optimal parameter vector, i.e, $w^*$, the training data set, i.e., $\mathcal D_{train}$ is need. The training data set is generated by a set of random simulated received power matrix based on the simulate reflection coefficient with specifically $\bm d^*$, i.e., $S(f, \bm d^*, \bm c)$ , the known distributions of environmental condition, i.e, $\mathcal C$ and the optimal \mm position set, i.e., $\mathcal X^*$, in the simulation environment.

After that, we optimize $f^w$ by training it on $\mathcal D_{train}$ using the supervised learning algorithm. The training of $\bm w$ is performed by iteratively updating $\bm w$ along the negative gradient the of the RMSE loss in~(\ref{sP3}), i.e.,
\beq
\setlength{\abovedisplayskip}{4pt}
\setlength{\belowdisplayskip}{4pt}
\bm w = \bm w - \beta \bigtriangledown_{\bm w} L_{RMSE}(\bm w, \mathcal X^*),
\eeq
where the gradient $\bigtriangledown_{\bm w} L_{RMSE}(\bm w, \mathcal X^*)$ is calculated by using the back-propagation algorithm and $\beta$ denotes the learning rate which used to control the updating rate. 

Sum up the algorithms to solve sub-problem (sP2) and (sP3), and we can summarize the complete algorithm to solve (P1) as Algorithm~\ref{Alg: system Optimization}.

\begin{algorithm}[t]  
  \caption{Algorithm to solve Meta-IoT system optimization.}  
  \label{Alg: system Optimization}
  \begin{algorithmic}[1]  
    \Require  
    $\mathcal S_X$ (available position set),  $S(f, \bm d^*, \bm c)$ (the simulate reflection coefficient with specifically $\bm d^*$), $\mathcal C$(known distribution of environmental condition), $\eta$, $R_{env}$, $N$;
    \Ensure  
    $\mathcal X^*$, $f^{w*}$;
    \State Solve the (sP2) and obtain the optimal \mm position set $\mathcal X^*$ by using the simulated annealing algorithm. 
    \State Based on $\mathcal C$, $\mathcal X^*$ and $S(f, \bm d^*, \bm c)$, generate the training data set $\mathcal D_{train}$.
    \State Based on $\mathcal X^*$, and $\mathcal D_{train}$ to train the neural network by solving (sP3), and obtain the optimized estimation function $f^{w*}$.
    \State \Return $X^*$, $f^{w*}$
   \end{algorithmic}  
\end{algorithm}  

\begin{table}[!t]
\vspace{-10pt}
\caption{Simulation Parameters}
\centering
\begin{small}
\begin{tabular}{|c|c|c|c|}
\hline
\textbf{Parameter}   &     \textbf{Value}  &  \textbf{Parameter} &   \textbf{Value}    \\
\hline
\makecell[c]{Frequency Range\\($[f_l, f_u]$)} &   \makecell[c]{3.5-4.5 \\GHz}    &  \makecell[c]{Number of samp-\\led frequencies ($L$)} &    101   \\
\hline
\makecell[c]{Number of SRR \\($N_s$)}    &      2       &      \makecell[c]{Number of decives \\($N$)}  &       10    \\
\hline
\makecell{Environment \\Dimension}&  \makecell{$5 \times 8 \times 3$\\$m^3$} & Space Grids ($M$)& $960$\\
\hline
\makecell{Device Area ($\sigma$)} & $0.01m^2$ & Antenna Array & $4 \times 4$\\
\hline
\makecell{Environment scatter\\ ratio ($\eta$)} & 0.9 & \makecell{Reflection coefficient\\ of wall ($R_{env}$)} & 0.5 \\
\hline
\end{tabular}
\end{small}
\label{T1}
\vspace{-5pt}
\end{table}

\section{Simulation Results} 
\label{sec: results} 
In this section, to validate the effectiveness of the proposed \mm system, we design and implement a \mm system to sense the temperature and humidity levels for in an indoor environment and present the simulation result. We first provide simulation results of the proposed system. Besides, we give insight into how the \mm devices' positions and quantity influence the system's precision.

In the simulation, \mm device consists of two SRRs. The first SRRs has temperature-sensitive material within its gap and is aimed for sensing temperature, which we refer to as the TSRR. Similarly, the second SRR contains humidity-sensitive material for sensing humidity and is referred to as HSRR. More specifically, the \mm is made of copper rings and FR-4 supportive substrate. The temperature-sensitive material is the powder used in NTC thermistor and the humidity-sensitive material is the polymer used in the hygristor. The \mm device's design is guided by the reflection coefficient model, which has proposed in Section~\ref{sec: mm sensor} and the gap widths is chosen as $\bm d^{*}$. Besides, the reflection coefficient, i.e., $S(f, \bm d^*, \bm c)$, is simulated with the help of CST software~\cite{CST}.

Besides, we denote the target space is a rectangular space and the \mm devices can be deployed on the wall. The transceivers are placed in the middle of space and the antenna array is composed with $4 \times 4$ omnidirectional antennas.
The detail simulation parameters are presented in Table \ref{T1}.

\begin{figure}[!t] 
    \center{\includegraphics[width=0.98\linewidth]{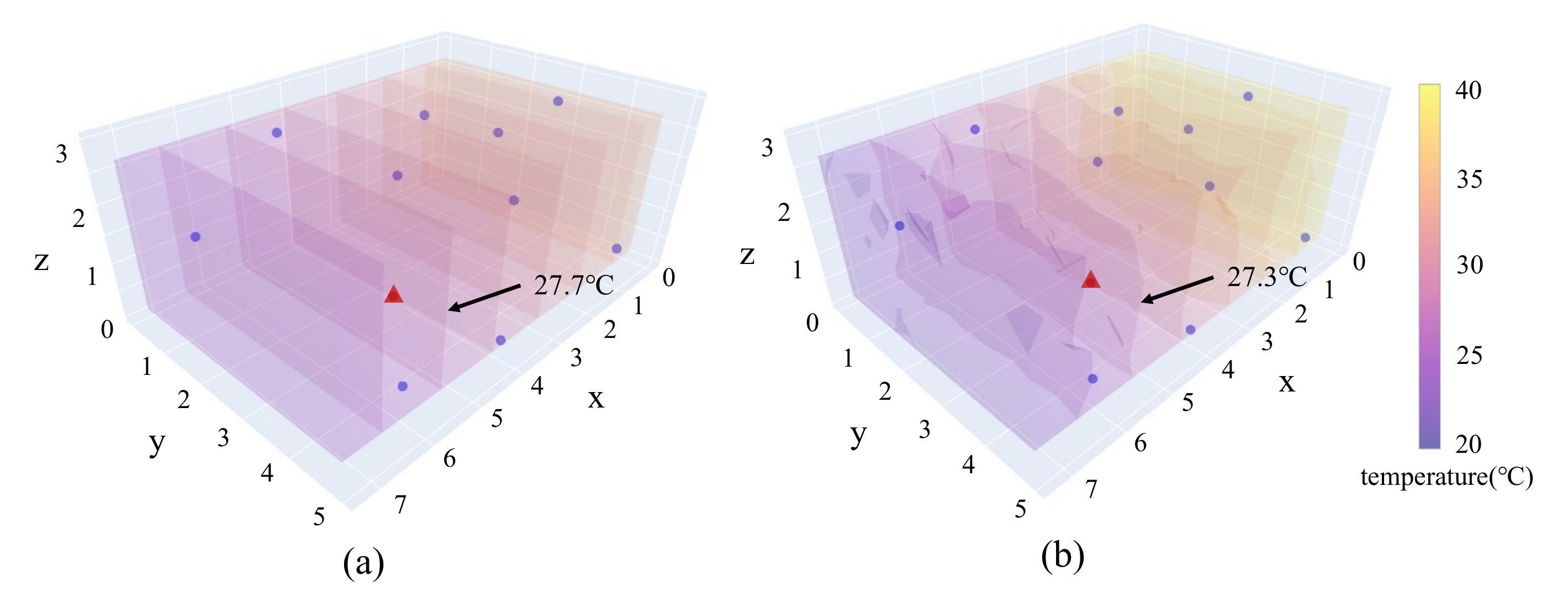}} 
	\caption{The reconstruction results for the test temperature distribution, (a) is the real distribution and (b) is the estimated distribution. In both figures, the red triangle represents the position of transceivers and the blue dots represent the \mm devices' position}
	\label{fig: reconstruct}
\end{figure}

Fig.~\ref{fig: reconstruct} shows the comparison between the reality distribution and the estimation   distribution for the testing temperature distribution by solving (P1). The test distribution is similar with the distributions, which are used to generate train set. Besides, in both figures, the red triangle represents the position of the transceivers and the blue dots represent the position of \mm devices, which has been optimized with (sP2). It can observe that the temperature gradient has been captured by the estimation function and the mean reconstruction error is less than $2.9^\circ$C.

\begin{figure}[!t] 
    \center{\includegraphics[width=0.9\linewidth]{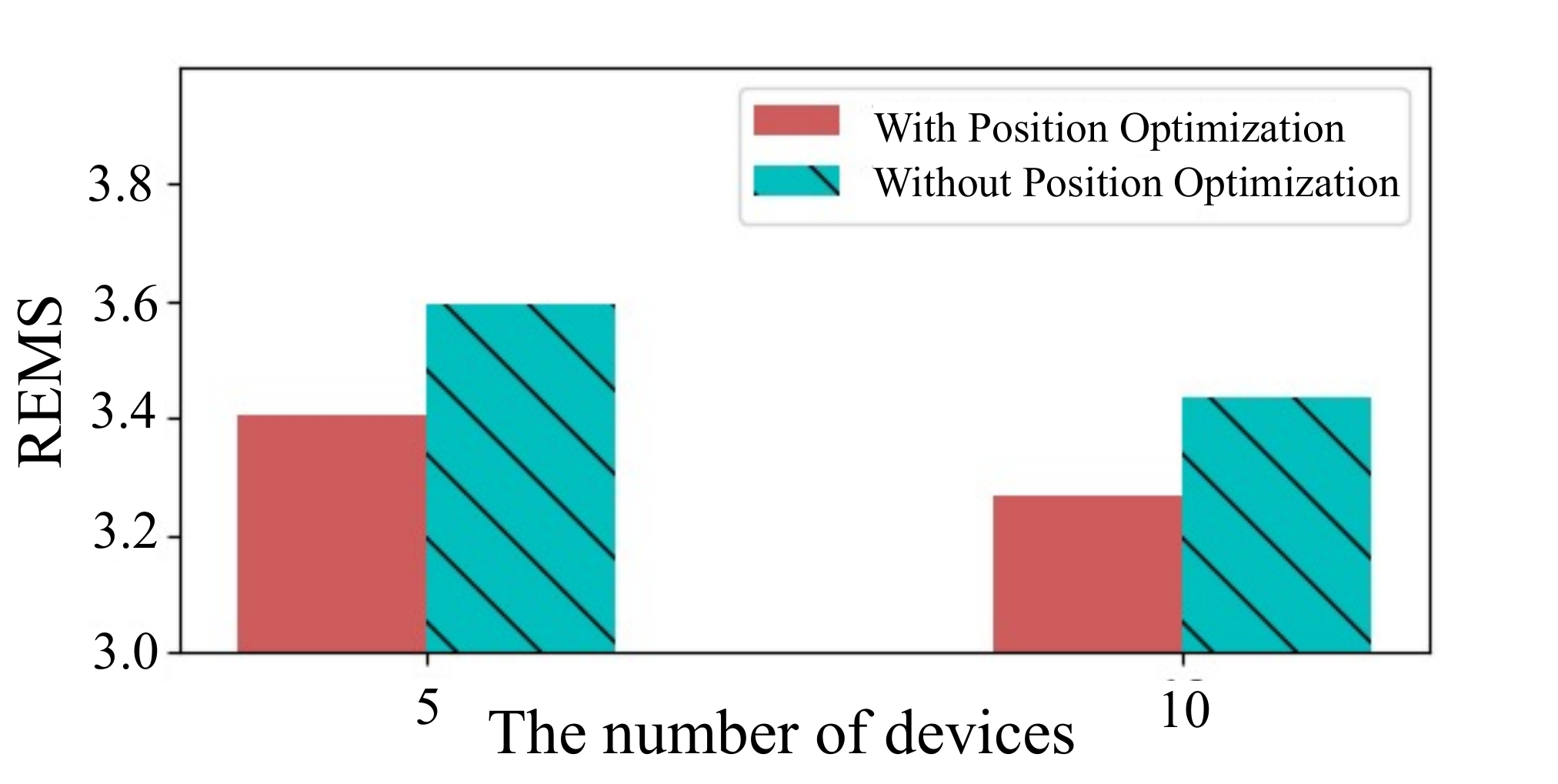}}
    \vspace{-0.3em} 
	\caption{Resulting RMSEs of the estimation function versus the number of deployed \mm devices, given different \mm position set.}
	\label{fig: loss and position}
	\vspace{-0.3em}
\end{figure}

Fig.~\ref{fig: loss and position} shows the different resulting RMSE by solve (sP3) under different \mm position-case. The first case indicates the resulting RMSE for the estimation function given the optimal position set ,i.e., $\mathcal X^*$, obtain by solve (sP2), and the second case indicates the resulting RMSE with random position set. Besides, there are five different distributions of temperature and humidity, and each of which has 256 generated datas to construct the training set, i.e., $\mathcal D_{train}$. It can be observed that the resulting RMSE values of different \mm position sets decrease as the number of deployed \mm device increase. Besides, the optimal \mm position leads to the lower RMSE under different number of device, which means the system can reach higher precision with the help of position optimization to minimize the interference.

\section{Conclusion}
\label{sec: conclusion}
In this paper, we have designed a novel technique to reconstruct the 3D distribution with \mm system. We have analyzed the relationship between \mm device's structure and its reflection coefficients as well as the transmission model. To handle the interference between \mm devices and minimize the reconstruction error, the interference between each \mm device has been considered and a joint position and estimation function optimization problem have been formulated and solved. Simulation results have shown the capability of the proposed system to reconstruct 3D distribution of temperature and humidity condition with the relative error less than~$9.7\%$.

\bibliographystyle{ieeetr}%
\bibliography{./bibilio.bib}
\end{document}